\documentclass[usenatbib]{mn2e}
\usepackage{epsfig}
\usepackage{amsmath}
\usepackage{ulem}
\usepackage{times}
\usepackage{longtable}
\usepackage{supertabular}

\def\gsim{\mathrel{\raise0.35ex\hbox{$\scriptstyle >$}\kern-0.6em 
\lower0.40ex\hbox{{$\scriptstyle \sim$}}}}
\def\lsim{\mathrel{\raise0.35ex\hbox{$\scriptstyle <$}\kern-0.6em 
\lower0.40ex\hbox{{$\scriptstyle \sim$}}}}

\date{\today}
\title[Dust extinction from H$\alpha$/UV ratio]
{Predicting dust extinction properties of star-forming galaxies \\ from H$\alpha$/UV ratio} 

\author[Y. Koyama et al.]{
\parbox[t]{\textwidth}{
Yusei Koyama,$^{\! 1,2}$\thanks{E-mail: koyamays@ir.isas.jaxa.jp}
Tadayuki Kodama,$^{\! 2}$
Masao Hayashi,$^{\! 2}$
Rhythm Shimakawa,$^{\! 3}$\\
Issei Yamamura,$^{\! 1}$
Fumi Egusa,$^{\! 1,2}$
Nagisa Oi,$^{\! 1}$
Ichi Tanaka,$^{\! 4}$
Ken-ichi Tadaki,$^{\! 5}$\\
Satoshi Takita,$^{\! 1}$
Sin'itirou Makiuti$^{\! 1}$
}
\vspace*{6pt}\\
$^{1}$Institute of Space Astronautical Science, Japan Aerospace Exploration Agency, Sagamihara, Kanagawa 252-5210, Japan\\
$^{2}$National Astronomical Observatory of Japan, Mitaka, Tokyo 181-8588, Japan\\
$^{3}$Department of Astronomical Science, The Graduate University for Advanced Studies, Mitaka, Tokyo 181-8588, Japan \\
$^{4}$Subaru Telescope, National Astronomical Observatory of Japan, 650 North A'ohoku Place, Hilo, HI 96720, USA\\
$^{5}$Max-Planck-Institut f{\"u}r Extraterrestrische Physik, Postfach 1312, Giessenbachstrasse, D-85741 Garching, Germany
}

\begin{document}

\maketitle

\begin{abstract}
Using star-forming galaxies sample in the nearby Universe (0.02$<$$z$$<$0.10) selected from the SDSS (DR7) and GALEX all-sky survey (GR5), we present a new empirical calibration for predicting dust extinction of galaxies from H$\alpha$-to-FUV flux ratio. We find that the H$\alpha$ dust extinction ($A_{\rm H\alpha}$) derived with H$\alpha$/H$\beta$ ratio (Balmer decrement) increases with increasing H$\alpha$/UV ratio as expected, but there remains a considerable scatter around the relation, which is largely dependent on stellar mass and/or H$\alpha$ equivalent width (EW$_{\rm H\alpha}$). At fixed H$\alpha$/UV ratio, galaxies with higher stellar mass (or galaxies with lower EW$_{\rm H\alpha}$) tend to be more highly obscured by dust. We quantify this trend and establish an empirical calibration for predicting $A_{\rm H\alpha}$ with a combination of H$\alpha$/UV ratio, stellar mass and EW$_{\rm H\alpha}$, with which we can successfully reduce the systematic uncertainties accompanying the simple H$\alpha$/UV approach by $\sim$15--30\%. The new recipes proposed in this study will provide a convenient tool for predicting dust extinction level of galaxies particularly when Balmer decrement is not available. By comparing $A_{\rm H\alpha}$ (derived with Balmer decrement) and $A_{\rm UV}$ (derived with IR/UV luminosity ratio) for a subsample of galaxies for which AKARI FIR photometry is available, we demonstrate that more massive galaxies tend to have higher extra extinction towards the nebular regions compared to the stellar continuum light. Considering recent studies reporting smaller extra extinction towards nebular regions for high-redshift galaxies, we argue that the dust geometry within high-redshift galaxies resemble more like low-mass galaxies in the nearby Universe.

\end{abstract}
\begin{keywords}
galaxies: evolution ---
galaxies: star formation ---
ISM: dust, extinction.

\end{keywords}
\section{Introduction}
\label{sec:intro}

Star formation rate (SFR) is one of the most fundamental parameters which characterise the nature of galaxies. Deriving SFRs of galaxies is therefore an important task in the extra-galactic astronomy. There are various indicators proposed for star-formation activity in galaxies, including rest-frame UV continuum light, nebular emission lines (such as H$\alpha$), mid- to far-infrared dust thermal emission, and radio continuum luminosity (see review by \citealt{kennicutt1998}; \citealt{kennicutt2012}). 

When measuring SFRs, dust extinction correction is always an important issue. Particularly in the case of star-burst galaxies, it is possible that only a tiny fraction of UV light can escape from the galaxy because they are heavily obscured by dust. This leads to an order of magnitude extinction correction for estimating the intrinsic UV luminosity (hence SFRs). For studies of high-redshift galaxies, where the cosmic star formation rate density is an order of magnitude higher than the present-day universe (e.g.\ \citealt{madau1996}; \citealt{hopkins2006}), luminous infrared galaxies (LIRGs) or ultra-luminous infrared galaxies (ULIRGs) are much more common population (e.g.\ \citealt{lefloch2005}; \citealt{magnelli2011}), and therefore the effect of dust must be carefully taken into account. Far-infrared (FIR) observations, on the other hand, allow a direct measurement of dust thermal emission, and in particular, observations covering a peak of spectral energy distribution (SED) of galaxies (usually located at around $\lambda_{\rm rest}$$\sim$100~$\mu$m) allow us to derive total IR luminosity (hence dust-enshrouded SFR) with reasonable accuracy. A problem is that such FIR observation requires space telescopes. Unfortunately, due to the poor spatial resolution of the IR space telescopes ever launched, the depths of FIR observations are always limited by source confusion, making it impossible to detect individual galaxies at high-redshift, except for exceptionally luminous objects.

Another important tool for measuring SFR is H$\alpha$$\lambda$6563 line, which is well-calibrated in the local Universe. The H$\alpha$ emission is emitted in star-forming H{\sc ii} regions (near short-lived, O-type stars), and so the H$\alpha$ luminosity is expected to be proportional to on-going SFR. The H$\alpha$ line is located at optical wavelength (hence less sensitive to dust extinction effects than UV light), and therefore H$\alpha$ line is recognised as an excellent indicator of SF activity. Of course, H$\alpha$ line is also affected by dust. Even for local spiral galaxies with moderate levels of star formation, the dust extinction at H$\alpha$ is not negligible ($A_{\rm H\alpha}\sim$1 mag: e.g.\ \citealt{kennicutt1983}). In the case of more active starbursts, the H$\alpha$ line is reported to be more heavily obscured by dust, and $A_{\rm H\alpha}$ exceeds $\sim$3-mag in extreme cases (\citealt{poggianti2000}; \citealt{koyama2010}). It is therefore ideally required to combine the H$\beta$$\lambda4861$ line flux, with which one can derive dust extinction level using H$\alpha$/H$\beta$ ratio (i.e. Balmer decrement). This method needs to assume the shape of extinction curve, as well as the electron density and temperature, but other than that, it allows us to derive dust extinction levels purely based on physics. With recent advents of sensitive NIR instruments installed on 8-m class telescopes, it is becoming easier to obtain H$\alpha$ information for high-redshift ($z\gsim 1$) galaxies; e.g.\ with spectroscopy and/or narrow-band imaging. However, H$\beta$ line is usually much fainter than H$\alpha$, making it very challenging to apply the Balmer decrement method for measuring dust extinction for high-$z$ galaxies. Indeed, it is becoming possible to obtain unprecedentedly high-quality NIR spectra for high-$z$ galaxies (e.g.\ \citealt{steidel2014}; \citealt{reddy2015}), but even with such high-quality data, it is still challenging to detect H$\beta$ lines from individual galaxies. 

Our idea is to use observed H$\alpha$-to-UV flux ratio to predict dust extinction levels of galaxies. For high-redshift studies, rest-frame UV flux density can easily be traced by optical photometry. Considering the fact that H$\alpha$ is less sensitive to dust extinction than UV continuum light, it is expected that the H$\alpha$/UV flux ratio can provide a crude test for dust extinction level (i.e.\ galaxies with higher H$\alpha$/UV ratio are expected to be dustier). It is of course true that an intrinsic $L_{\rm H\alpha}$/$L_{\rm UV}$ ratio can change with galactic age, SF history, metallicity, or initial mass function (IMF), and therefore it is not straightforward to directly link the H$\alpha$/UV ratio to the dust extinction properties (\citealt{wuyts2013}; \citealt{lee2009}; \citealt{pflamm2009}; \citealt{zeimann2014}). Nevertheless, as demonstrated by \cite{buat2002} with a small sample of local star-forming galaxies, the H$\alpha$/UV ratio is indeed well correlated with dust extinction properties. In this paper, we will revisit this issue and attempt to construct an empirical calibration to predict dust extinction with H$\alpha$/UV ratio, by compiling a statistical sample of nearby galaxies ($z<0.1$) drawn from SDSS, GALEX, and AKARI.  

This paper is organised as follows. In Section~2, we present our datasets and summarize physical quantities used in the paper. We use SDSS, GALEX, and AKARI data to obtain UV-, H$\alpha$-, and FIR-based SFRs. Our main results are shown in Section~3. We first present a positive correlation between dust extinction ($A_{\rm H\alpha}$) and H$\alpha$/UV flux ratio (Section~3.1), and then we draw the scatter around the $A_{\rm H\alpha}$--H$\alpha$/UV correlation as functions of stellar mass and/or H$\alpha$ equivalent width (Section~3.2--3.4). We establish an useful prescription for predicting dust extinction in the absence of H$\beta$ line or FIR photometry (Section~3.5). In Section~4, we present related analyses and discuss the dust properties of star-forming galaxies. We first provide a careful investigation of the systematic uncertainties associated with aperture correction for SDSS data (Section~4.1). We then discuss if we can really derive the dust extinction levels of galaxies by using stellar mass alone (Section~4.2). We also discuss the relation between dust extinction and metallicity in Section~4.3. Furthermore, in Section~4.4, we discuss the extra extinction towards nebular regions, and report its dependence on stellar mass and EW$_{\rm H\alpha}$. Finally, we will discuss if our new recipes can be applicable to high-$z$ galaxies using our high-$z$ galaxy sample (Section~4.5). Our conclusion is given in Section~5. Throughout the paper, we adopt $\Omega_{\rm{M}} =0.3$, $\Omega_{\Lambda} =0.7$, and $H_0 =70$~km~s$^{-1}$Mpc$^{-1}$.

\begin{figure}
\vspace{-5mm}
\hspace{-10mm} 
\includegraphics[angle=0,width=10.7cm]{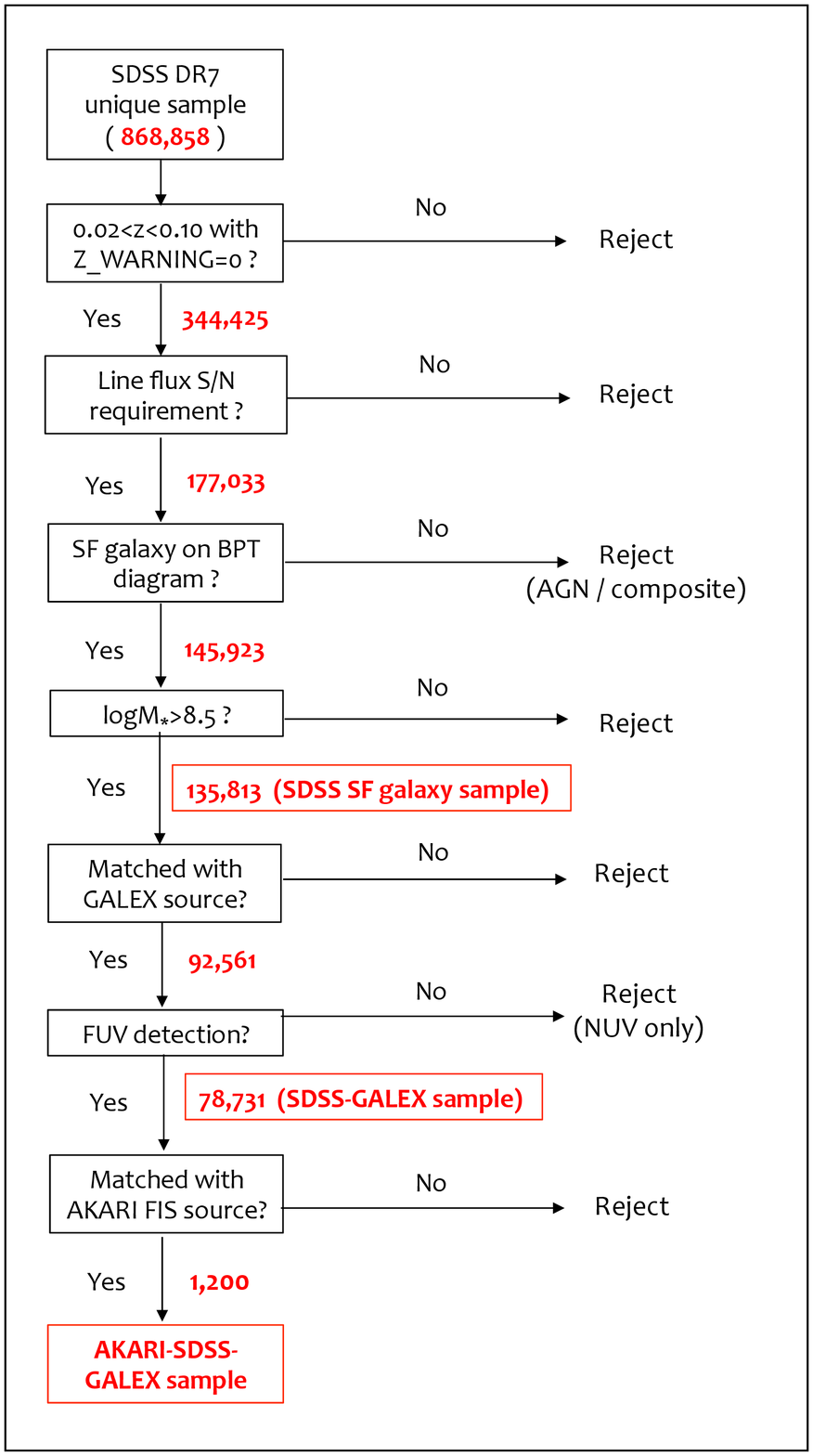} 
\vspace{-1.0cm}
\caption{Summary of our sample selection and the number of the galaxy sample at each step during the cross-identification between SDSS, GALEX, and AKARI sources.}
\label{fig:flow_chart}
\end{figure}
\begin{figure*}
\vspace{0mm}
 \begin{center}
  \includegraphics[angle=0,width=14.7cm]{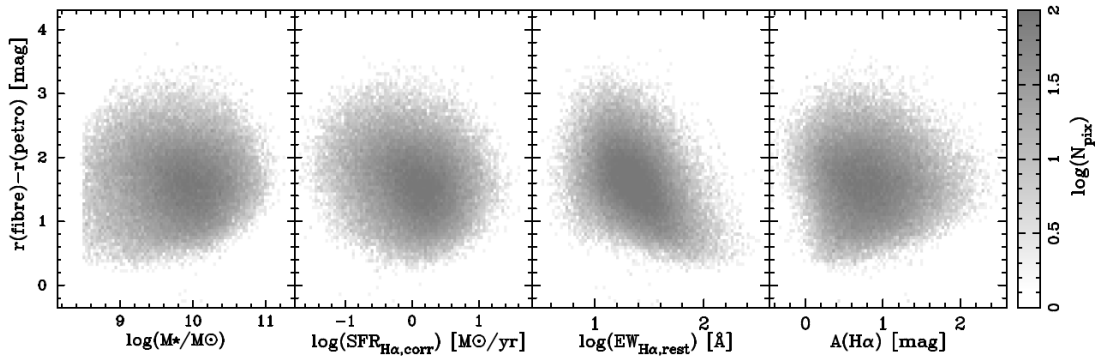} 
 \end{center}
\vspace{-2mm}
\caption{Difference between total (petrosian) magnitudes and SDSS fibre magnitudes at $r$-band as functions of stellar mass, SFR$_{\rm H\alpha}$, EW$_{\rm H\alpha}$, and $A_{\rm H\alpha}$ (from left to right). There seems to be a decreasing trend of the magnitude difference with EW$_{\rm H\alpha}$, which will be explained by the fact that high-EW galaxies tend to be more compact, whilst the correlation between the magnitude difference and other galaxy properties are much weaker. To create this plot, we divide each panel into 90$\times$90 sub-grid, and count the number of galaxies in each pixel. Note that we basically follow this strategy when we show the grey-scale or colour image plot in the remainder of this paper, unless otherwise stated.}
\label{fig:aperture_corr1}
\end{figure*}

\section{Data and sample selection}
\label{sec:data}

\subsection{SDSS data}

We use the spectroscopic catalogue of the Sloan Digital Sky Survey (SDSS) Data Release 7 (DR7; \citealt{abazajian2009}). The spectroscopic measurements are performed by the Max Planck Institute for Astrophysics and Johns Hopkins University (MPA/JHU group), and we make use of the ``value-added'' catalogues retrieved from their public website\footnote{http://www.mpa-garching.mpg.de/SDSS/DR7/}. The catalogue contains a total of 927,552 objects, of which 868,858 are identified as unique sources: we here perform internal matching using 2$''$ search radius to avoid duplicated objects. Furthermore, we restrict the sample with a redshift range of 0.02$<$$z$$<$0.10, and exclude those having uncertain redshifts (by applying {\sc z\_warning}$=$0). With these criteria, we have selected 344,425 objects. The redshift range adopted here was chosen so that we can study galaxies in a wide range in stellar mass, and at the same time we can minimise a potential effect of redshift evolution of galaxy properties within the sample. We note that a small change of the redshift range applied here does not affect our results.

We select star-forming galaxies using the 'BPT' diagram (\citealt{baldwin1981}). The BPT diagnostics requires [O{\sc iii}]$\lambda$5007/H$\beta$ and [N{\sc ii}]$\lambda$6584/H$\alpha$ line flux ratio, and therefore these four major emission lines need to be detected. We first request S/N(H$\alpha$)$>$10 (as H$\alpha$ line is usually the strongest), yielding 196,073 sample. We also request the following criteria for S/N ratio of the other three emission lines: S/N(H$\beta$)$>$3, S/N([N{\sc ii}])$>$3, S/N([O{\sc iii}])$>$2. We note that [O{\sc iii}] line is the weakest in most cases, and so we accept the [O{\sc iii}] line detection down to S/N$>$2. After applying all these criteria, 177,033 galaxies are selected (i.e.\ $>$90\% of the H$\alpha$-selected galaxies (with S/N$>$10) have significant detection at all the other three lines). Following the recommendation by the MPA/JHU group, we scaled the uncertainty of H$\alpha$, H$\beta$, [N{\sc ii}], and [O{\sc iii}] by 2.473, 1.882, 2.039, and 1.566, respectively. We also note that the line flux measurements provided in the MPA/JHU catalogue are continuum-subtracted (and also corrected for Galactic reddening), and so the effect of stellar absorption is properly taken into account. 

Following the prescription by \cite{kauffmann2003b}, we select SF galaxies with: 
\begin{equation}
\log {\rm [OIII]/H\beta} < 0.61/(\log({\rm [NII]/H\alpha})-0.05) + 1.3,
\end{equation} 
and we also apply [N{\sc ii}]/H$\alpha$$<$0.6 to remove any Seyfert galaxies or LINERs (\citealt{kauffmann2003b}; \citealt{kewley2006}). We have now selected 145,923 {\it star-forming} galaxies. 

Our final requirement for the sample selection is the availability of stellar mass estimates ($M_{\star}$). Stellar mass of the SDSS DR7 galaxies are computed by the MPA/JHU group, by fitting to the SDSS broad-band photometry following the philosophy of \cite{kauffmann2003a} and \cite{salim2007}. In this work, we use galaxies with $\log (M_{\star}/M_{\odot})$$>$8.5. Our results do not change even if we do not apply the stellar mass cut, but we note that the sample size significantly decreases below $\log (M_{\star}/M_{\odot})$$=$8.5 (by a factor of $\sim$20$\times$ compared with those at the peak of the $M_{\star}$ distribution), and therefore we decided not to use galaxies below $\log (M_{\star}/M_{\odot})$$=$8.5 for statistical analyses presented in this work. Overall, our final SF galaxy sample contains 135,813 galaxies (with the median redshift of $z=0.063$). Our sample selection procedure described here is also outlined in Fig.~\ref{fig:flow_chart}.

We derive H$\alpha$ dust extinction ($A_{\rm H\alpha}$) for each galaxy using H$\alpha$/H$\beta$ flux ratio (i.e.\ Balmer decrement) using the following equation:  
\begin{equation}
A_{\rm H\alpha}=\frac{-2.5k_{\rm H\alpha}}{k_{\rm H\alpha}-k_{\rm H\beta}} \log \left( \frac{2.86}{F_{\rm H\alpha}/F_{\rm H\beta}} \right),
\end{equation}
where the term 2.86 is the intrinsic H$\alpha$/H$\beta$ flux ratio for Case B recombination at a temperature of 10$^4$ K and at an electron density of $n_{\rm e}$$=$10$^2$~cm$^{-3}$. Following \cite{garn2010b}, we assume \cite{calzetti2000} dust extinction law to calculate $k_{\rm H\alpha}$ and $k_{\rm H\beta}$, so that $A_{\rm H\alpha}$ is derived in a more specific form of:
\begin{equation}
A_{\rm H\alpha} = 6.53 \log\left( F_{\rm H\alpha}/F_{\rm H\beta} \right) - 2.98,  
\end{equation}
with the median uncertainty of $\sim$0.24 mag (based on the uncertainties regarding the H$\alpha$/H$\beta$ measurements). We then estimate star formation rate (SFR) of each galaxy based on its H$\alpha$ luminosity. After correcting the dust extinction using the $A_{\rm H\alpha}$ value derived above, we compute SFR$_{\rm H\alpha}$ using the standard calibration of \cite{kennicutt1998} for \cite{kroupa2001} IMF\footnote{\cite{kennicutt2009} note that the zero point of the SFR needs to be reduced by a factor of 1.44 in the case of Kroupa IMF, compared with the original \cite{kennicutt1998} calibration (which assumes \cite{salpeter1955} IMF).}: SFR$_{\rm H\alpha}$$=$5.5$\times$10$^{-42}$$L_{\rm H\alpha}$ [erg\,s$^{-1}$]. We chose Kroupa IMF to be consistent with stellar mass estimates for SDSS galaxies performed by MPA/JHU group (\citealt{kauffmann2003a}), but the choice of IMF does not affect our conclusion as our results are mostly based on the {\it ratio} of SFR$_{\rm H\alpha}$ and SFR$_{\rm UV}$. 

We note that SDSS spectroscopic measurements were performed with a limited size of fibre (3-arcsec in diameter). We therefore apply an aperture correction based on the difference between total and fibre-based $r$-band magnitudes: i.e. we assume that the emission-line profile follows that of continuum light. We believe that this assumption is reasonable, as supported by previous studies (\citealt{brinchmann2004}; \citealt{kewley2005}), and using our own sample, we also check the difference between fibre- and total-magnitudes (at $r$-band) as functions of various galaxy properties (Fig.~\ref{fig:aperture_corr1}). We find that the fibre magnitudes sometimes trace only $\sim$10\% of total light for a fraction of galaxies, but we also find that the level of aperture correction does not strongly correlate with physical properties ($M_{\star}$, SFR, $A_{\rm H\alpha}$). An exception is that there is a decreasing trend of the magnitude difference with EW$_{\rm H\alpha}$, which is probably because galaxies having higher EW$_{\rm H\alpha}$ tend to be more centrally concentrated. We also verify that our aperture-corrected SFR$_{\rm H\alpha}$ show good agreement with those derived from AKARI FIR photometry (see also Section~2.3). Thus, we believe that the aperture correction does not make any artificial trend, but the results presented in this work need to be verified independently by e.g. using large aperture photometry for measuring H$\alpha$ emission\footnote{We speculate that the typical uncertainties regarding the aperture correction would be $\sim$0.15~dex (or $\sim$0.4~mag), by computing the standard deviation of the difference between the aperture correction values measured at $u$-band and $r$-band (i.e.\ $\Delta u - \Delta r$: see also Section~4.1). However, it is not straightforward to determine the uncertainties associated with the aperture correction, and we do not consider uncertainties regarding aperture correction in the remainder of this work.}. More detailed discussion regarding the aperture correction will be provided in Section.~4.1.

\begin{figure}
\vspace{2mm}
 \begin{center}
  \includegraphics[angle=0,width=7.0cm]{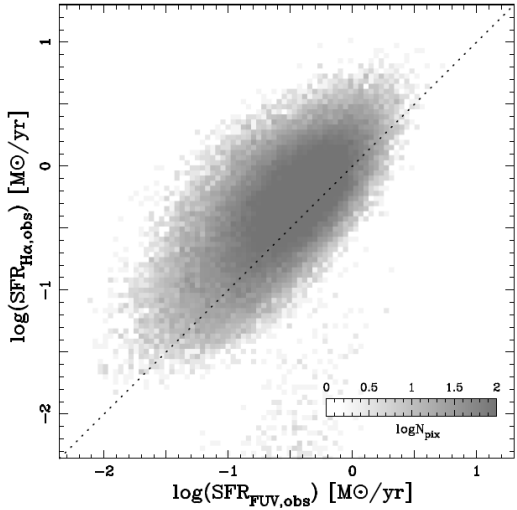}
 \end{center}
\vspace{-2mm}
\caption{Comparison between the observed H$\alpha$-derived SFRs (with aperture correction) and UV-based SFRs for our SDSS--GALEX sample. The dotted line shows the one-to-one correlation; SFR$_{\rm H\alpha,obs}$$=$SFR$_{\rm FUV,obs}$. A systematic offset towards higher SFR$_{\rm H\alpha}$ reflects the different level of dust extinction for H$\alpha$ and UV (i.e.\ H$\alpha$ is less sensitive to dust extinction). }
\label{fig:SFR_comparison_UV}
\end{figure}

\subsection{GALEX data}

The goal of this paper is to draw dust extinction properties of galaxies as a function of H$\alpha$/UV ratio. We here perform cross-identification between our SDSS galaxy sample with Galaxy Evolution Explorer ({\it GALEX}; \citealt{martin2005}) UV sources. We use unique GALEX sources selected from All-sky Imaging Survey (AIS) of GALEX fifth data release (GR5), reaching down to the depth of $\sim$19.9 and $\sim$20.8 [AB mag] at far-UV (FUV; $\lambda_{\rm eff}$$=$1516~\AA) and near-UV (NUV; $\lambda_{\rm eff}$$=$2267~\AA), respectively. Here we use the catalogue published by \cite{bianchi2011}, available on the MAST web site\footnote{http://archive.stsci.edu/prepds/bianchi\_gr5xdr7/}. We cross-identified our SDSS (spectroscopic) galaxy sample with GALEX sources using 3$''$ search radius. We find that 92,561 SF galaxies (out of 135,813 galaxies selected in Section~2.1) have UV counterparts, among which 78,731 galaxies are detected at both FUV and NUV (because of the limited depth at FUV). In this paper, we use star-forming galaxies detected at both FUV and NUV as the ``SDSS--GALEX'' sample. 

We apply the Galactic extinction corrections for the UV photometry using the \cite{schlegel1998} dust map and the Galactic extinction curve of \cite{cardelli1989} for $R_V=3.1$: specifically, we apply $A_{\rm FUV}=7.9E(B-V)$ and $A_{\rm NUV}=8.0E(B-V)$. We also apply $k$-correction for the UV photometry using publicly available $k$-correction tool (\citealt{chilingarian2010}; 2012). We use FUV$-$NUV and NUV$-$$r$ colours to predict $k$-correction of individual galaxies at FUV and NUV, respectively. In this paper, we use FUV photometry to compute UV-based SFRs, because FUV photometry is expected to be more reliable than NUV because fluxes at NUV wavelength are often contributed by stars with life time of $>$100~Myr. We note that the typical $k$-correction value at FUV band turns out to be only $\sim$0.02--0.03~mag level, and so the effect on our results is negligible. 

We derive SFR$_{\rm UV}$ following the standard \cite{kennicutt1998} calibration assuming the Kroupa IMF to be consistent with the H$\alpha$ SFRs:
SFR$_{\rm UV}$$=$9.7$\times$10$^{-29}L_{\nu,{\rm FUV}}$~[erg\,s$^{-1}$\,Hz$^{-1}$]. We stress again that the choice of IMF would not strongly affect our results since we discuss the ratio of H$\alpha$- and UV-based SFRs (and the focus of this paper is mainly on massive star formation where most IMFs agree). In Fig.~\ref{fig:SFR_comparison_UV}, we compare the observed SFR$_{\rm H\alpha}$ and SFR$_{\rm FUV}$ (i.e.\ without dust extinction correction for both SFRs). There is a general trend that SFR$_{\rm H\alpha}$ tend to be higher than SFR$_{\rm FUV}$, which is likely because of the dust extinction effect, as will be discussed later in this paper. 

\begin{figure}
\vspace{2mm}
 \begin{center}
  \includegraphics[angle=270,width=7.0cm]{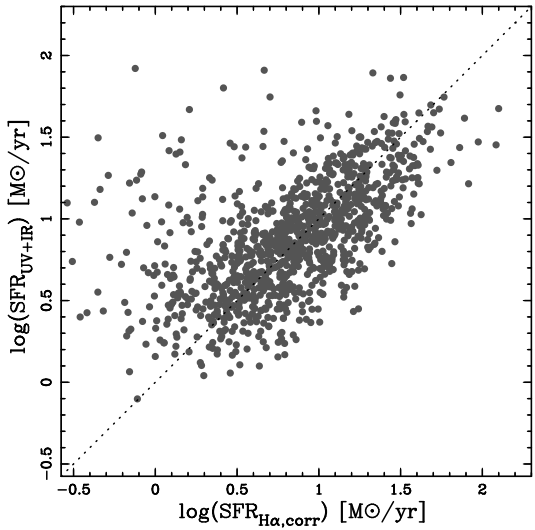} 
 \end{center}
\vspace{-2mm}
\caption{SFR from UV$+$IR photometry plotted against H$\alpha$-based SFRs (with aperture and dust extinction correction) for all SF galaxies in our AKARI--SDSS--GALEX sample, demonstrating a good agreement between the two independent measurements of SFR, and supporting our procedure for H$\alpha$ aperture and dust extinction correction works reasonably well.}
\label{fig:SFR_comparison_IR}
\end{figure}

\subsection{AKARI data}

We also match our SDSS--GALEX sample with the AKARI FIS bright source catalogue (ver.~1; \citealt{yamamura2009}; 2010) to obtain FIR fluxes for individual galaxies. The AKARI satellite (\citealt{murakami2007}) is a Japanese infrared space telescope, which performed all-sky survey in the MIR and FIR wavelength range (\citealt{ishihara2010}; \citealt{doi2015}; \citealt{takita2015}). Sources listed in the AKARI FIS bright source catalogue is selected and flux-limited at 90$\mu$m. The catalogue includes photometric information at four FIR bands (60, 90, 140, 160$\mu$m). Their 5$\sigma$ sensitivity at each band is 2.4, 0.55, 1.4, and 6.3 Jy, for N60(60$\mu$m), Wide-S(90$\mu$m), Wide-L(140$\mu$m), and N160(160$\mu$m), respectively. Most of the AKARI FIR sources are expected to be Galactic objects, but it also contains a substantial number of extra-galactic sources (see e.g.\ \citealt{pollo2010}; \citealt{goto2011a}b; \citealt{yuan2012}; \citealt{toba2013}; \citealt{totani2014}; \citealt{kilerci2014}). Considering the PSF size of $\sim$40 arcsec of the AKARI 90$\mu$m data, we search counterparts for the AKARI FIS sources within our SDSS--GALEX catalogue (i.e.\ 78,731 star-forming galaxies at $0.02<z<0.1$; see Section~2.2) using 20$''$ radius, and find 1200 AKARI/FIS sources have counterparts in the SDSS--GALEX catalogue with the median redshift of $z$$=$0.038. In this work, we use this 1200 SF galaxies as our final ``AKARI--SDSS--GALEX'' sample. 

Because of the limited depths of AKARI all-sky survey, most galaxies in the AKARI--SDSS--GALEX catalogue are not detected at all four AKARI FIS bands. We therefore derive total infrared luminosity ($L_{\rm IR}$) using the WIDE-S (90$\mu$m) and WIDE-L (140$\mu$m) photometry as presented by \cite{takeuchi2010}: 
\begin{equation}
\log L_{\rm IR} = 0.964\log L^{\rm 2band}_{\rm AKARI} + 0.814,
\end{equation}
where $L^{\rm 2band}_{\rm AKARI} = \Delta \nu_{\rm 90\mu m} L_{\nu}({\rm 90\mu m}) + \Delta \nu_{\rm 140\mu m} L_{\nu}({\rm 140\mu m})$. We note that $\Delta \nu _{\rm 90\mu m} = 1.47\times 10^{12}$[Hz] and $\Delta \nu _{\rm 140\mu m}=0.831\times 10^{12}$[Hz] denote the band width for the AKARI WIDE-S and WIDE-L band, respectively (see also \citealt{hirashita2008}). We derive 90$\mu$m and 140$\mu$m luminosity density from the observed flux density by multiplying $4\pi d_L^2 / (1+z)$ for each galaxy\footnote{We note that 1117 out of 1200 galaxies ($\sim$94\%) within our AKARI--SDSS--GALEX sample have photometry at both 90$\mu$m and 140$\mu$m. The remaining 73 galaxies ($\sim$6\%) are detected only at 90$\mu$m. We do not estimate $L_{\rm IR}$ for those detected only at 90$\mu$m, and they are not used in the following analyses.}. Here we do not consider $k$-correction term as the AKARI/FIS band widths are wide enough, and its effect is negligible for our low-redshift galaxy sample ($z<0.1$). 

The IR-based SFRs (SFR$_{\rm IR}$) are derived using \cite{kennicutt1998} equation with the Kroupa IMF to be consistent with our UV- and H$\alpha$-based SFRs; i.e.\ SFR$_{\rm IR}$$=$3.1$\times$$10^{-44}L_{\rm IR}$ [erg$\cdot$s$^{-1}$]. In Fig.~\ref{fig:SFR_comparison_IR}, we compare SFR$_{\rm UV+IR}$($=$SFR$_{\rm UV}+$SFR$_{\rm IR}$) and dust- and aperture-corrected H$\alpha$-based SFRs. The scatter is not small ($\sim$0.3~dex), but these two completely independent measurements show a good agreement (with a median difference of $\sim$0.01~dex), considering the uncertainties associated with AKARI fluxes, as well as the H$\alpha$ aperture/dust correction. We note that the catalogue constructed here will be published on the AKARI website.

\begin{figure}
\vspace{0mm}
 \begin{center}
  \includegraphics[angle=0,width=7.5cm]{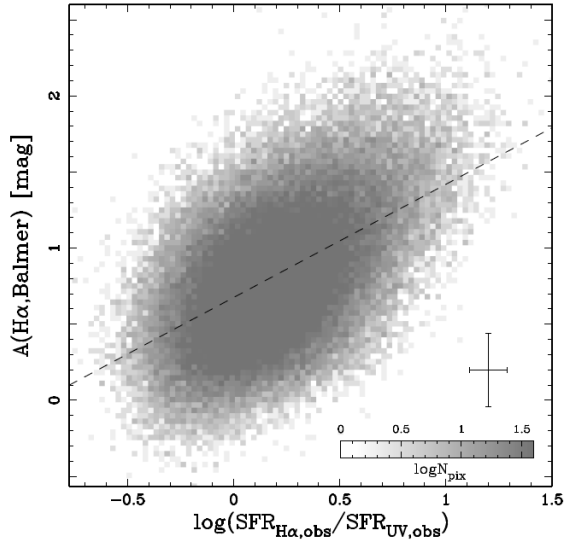} 
 \end{center}
\vspace{-3mm}
\caption{The H$\alpha$ extinction ($A_{\rm H\alpha}$) derived from Balmer decrement plotted against the observed SFR$_{\rm H\alpha}$/SFR$_{\rm UV}$ ratio for all SF galaxies in our SDSS--GALEX sample. Dust extinction correction is not applied when computing SFR$_{\rm H\alpha,obs}$/SFR$_{\rm UV,obs}$. The dashed line shows the best-fitted relation computed using the galaxies within the range of $-$0.5$<$$\log$(H$\alpha$/UV)$<$1.5. There exists a positive correlation between the two quantities (as expected), but there remains a considerable scatter around the best-fitted relation. In the bottom-right corner of this plot, we also show the typical (median) error bars for individual data points derived from the flux uncertainties.}
\label{fig:AHa_vs_HaUV}
\end{figure}
\begin{figure*}
\vspace{1mm}
 \begin{center}
 \includegraphics[angle=0,width=16.5cm]{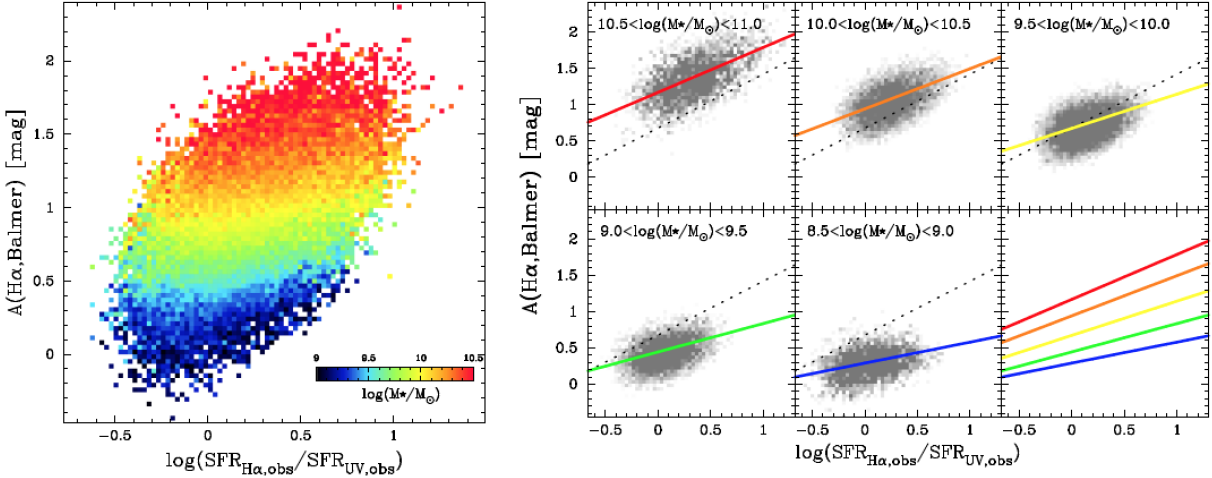} 
 \end{center}
\vspace{-3mm}
\caption{({\it Left}): The $A_{\rm H\alpha}$ versus H$\alpha$/UV plot (same as Fig.~\ref{fig:AHa_vs_HaUV}), colour-coded according to the average stellar mass at each position. We here applied 90$\times$90 gridding, and compute average $\log(M_{\star}/M_{\odot})$ in each grid. We note that we request a minimum sample size of $N_{\rm pix}$$=$4 for computing average stellar mass. We note that we apply this strategy when making similar plots in the remainder of the paper. ({\it Right}): The $A_{\rm H\alpha}$ versus SFR$_{\rm H\alpha}$/SFR$_{\rm UV}$ diagram for different stellar mass bin as indicated in the plot, demonstrating that dust extinction level is strongly dependent on stellar mass of galaxies even at fixed H$\alpha$/UV ratio, in the sense that more massive galaxies are more highly obscured by dust. We apply 60$\times$60 gridding (instead of 90$\times$90) to create this plot, because of the smaller sample size in each sub-panel. The dotted line indicates the best-fitted relation for the total sample (Eq.~5).}
\label{fig:AHa_vs_HaUV_mass}
\end{figure*}

\section{Results}

\subsection{Simple conversion from H$\alpha$/UV ratio to $A_{\rm H\alpha}$}

Our first step is to check if there is really a positive correlation between the $A_{\rm H\alpha}$ and H$\alpha$/UV ratio, using all our star-forming galaxy sample. The H$\alpha$ luminosities and FUV luminosity density are both proportional to recent SF activity, whereas these two indicators are affected by dust extinction at a different level. Because H$\alpha$ line is less sensitive to dust extinction effects, galaxies with higher H$\alpha$/UV ratio are expected to be dustier. We plot in Fig.~\ref{fig:AHa_vs_HaUV} the $A_{\rm H\alpha}$ (derived from Balmer decrement) of all SF galaxies in our SDSS--GALEX sample against their SFR$_{\rm H\alpha, obs}$/SFR$_{\rm FUV, obs}$ ratio\footnote{We note that all the analyses presented in this paper are based on the SFR$_{\rm H\alpha, obs}$/SFR$_{\rm FUV, obs}$ ratio for convenience, but our results are unchanged even if we instead use e.g. $L_{\rm H\alpha, obs}$/$\nu L_{\rm FUV, obs}$ (without converting their luminosities to SFRs).}. As expected, there exists a positive correlation between $A_{\rm H\alpha}$ and SFR$_{\rm H\alpha, obs}$/SFR$_{\rm FUV, obs}$ with the best-fitted relation of:
\begin{equation}
A_{\rm H\alpha} = 0.743 \times \log ({\rm H\alpha/UV}) + 0.676,
\end{equation}
where H$\alpha$/UV denotes the observed H$\alpha$-to-FUV SFR ratio (SFR$_{\rm H\alpha, obs}$/SFR$_{\rm FUV, obs}$). At the same time, it is also notable that there exists a substantial scatter around the best-fitted relation ($\sigma$$\sim$0.4~mag). Therefore, the data suggest that the dust extinction level of galaxies can (roughly) be approximated by H$\alpha$/UV ratio, but it may be too uncertain to derive dust extinction by simply applying the best-fitted $A_{\rm H\alpha}$--H$\alpha$/UV relation shown as the dashed line in Fig.~\ref{fig:AHa_vs_HaUV}. In the remainder of this paper, we will examine the origin of this scatter, and attempt to reduce the systematic uncertainty in deriving the dust extinction from UV--optical information. 

 In Fig.~\ref{fig:AHa_vs_HaUV}, a fraction of galaxies show $\log$(H$\alpha$/UV)$<$0 (i.e.\ SFR$_{\rm H\alpha}$$<$SFR$_{\rm UV}$). One may find it surprising, because it is expected that SFR$_{\rm H\alpha, obs}$ must be higher than SFR$_{\rm FUV, obs}$ under the assumption that H$\alpha$ is less sensitive to dust extinction ($A_{\rm FUV}$$>$$A_{\rm H\alpha}$). However, as discussed by e.g.\ \cite{lee2009}, UV-based SFR can exceed H$\alpha$-based SFR for various reasons, particularly for low-mass galaxies. Indeed, most of the galaxies exhibiting $\log$(H$\alpha$/UV)$<$0 in our sample turned out to be low-mass galaxies (see Section~4.2). We emphasize that the aim of this paper is to empirically link the observed H$\alpha$/UV flux ratio to the H$\alpha$/H$\beta$ line ratio (hence dust extinction level), and so we do not discuss this issue further in detail in the remainder of this paper.

\subsection{Dependence on stellar mass}

Recent studies have shown that dust extinction of galaxies are strongly correlated with stellar mass (e.g.\ \citealt{garn2010b}; \citealt{reddy2010}). Therefore we first investigate stellar mass dependence of the $A_{\rm H\alpha}$--H$\alpha$/UV relation. In the left panel of Fig.~\ref{fig:AHa_vs_HaUV_mass}, we show how the average stellar mass of galaxies change on the A$_{\rm H\alpha}$--H$\alpha$/UV scattered plot. Fig.~\ref{fig:AHa_vs_HaUV_mass} is the same plot as Fig.~\ref{fig:AHa_vs_HaUV}, but colour-coded according to the average stellar mass computed at each point (redder colours indicate higher $M_{\star}$). We here divide our sample into 90$\times$90 grid and compute average stellar mass $\langle \log(M_{\star}/M_{\odot}) \rangle$ in each grid. We request a minimum sample size of $N_{\rm pix}=4$ for computing the average stellar mass, and pixels containing the smaller number of galaxies (i.e. with $N_{\rm pix}\le 3$) are not shown. 

A visual inspection of the left panel of Fig.~\ref{fig:AHa_vs_HaUV_mass} reveals that more massive galaxies tend to be more highly obscured by dust at fixed H$\alpha$/UV ratio. By comparing the most massive galaxies ($\log(M_{\star}/M_{\odot})\gsim 10.5$) and low-mass galaxies ($\log(M_{\star}/M_{\odot}) \lsim 9.0$) in our sample, we find that the systematic difference in terms of $A_{\rm H\alpha}$ at fixed H$\alpha$/UV ratio can reach $\gsim$1.5-mag level at maximum. This result suggests that it might be misleading to blindly convert H$\alpha$/UV ratio into $A_{\rm H\alpha}$ for individual galaxies without considering the stellar mass difference. In the right panel of Fig.~\ref{fig:AHa_vs_HaUV_mass}, we plot $A_{\rm H\alpha}$ against H$\alpha$/UV ratio by dividing the sample into five stellar mass bins. This plot also demonstrates that the location of galaxies on the $A_{\rm H\alpha}$--H$\alpha$/UV diagram is largely dependent on their stellar mass: high-mass and low-mass galaxies dominate completely different regions on this diagram. 

It is also interesting to point out that the ``slope'' of the $A_{\rm H\alpha}$--H$\alpha$/UV relation can change with stellar mass. The non-zero slope of the $A_{\rm H\alpha}$--H$\alpha$/UV relation at fixed stellar mass seen in Fig.~\ref{fig:AHa_vs_HaUV_mass} implies that the dust extinction cannot be determined solely from stellar mass alone, and this result motivates us to establish an empirical calibration for deriving more realistic estimates of $A_{\rm H\alpha}$ using the H$\alpha$/UV ratio {\it and} stellar mass of galaxies. 

We here assume the following conversion equation: 
\begin{eqnarray}
A_{\rm H\alpha} = a(\log M_{\star})\times \log ({\rm H\alpha/UV}) + b(\log M_{\star}),
\end{eqnarray}
where $a(\log M_{\star})= a_1\log M_{\star} + a_2$ and $b(\log M_{\star})= b_1\log M_{\star} + b_2$. We fit all SDSS--GALEX sample with the above equation, and the resultant $a(\log M_{\star})$ and $b(\log M_{\star})$ are derived as follows:
\begin{eqnarray}
a(\log M_{\star}) = 0.210\times \log M_{\star} -1.597, \\
b(\log M_{\star}) = 0.493\times \log M_{\star} -4.121.
\end{eqnarray}
\begin{figure*}
\vspace{1mm}
 \begin{center}
   \includegraphics[angle=0,width=16.5cm]{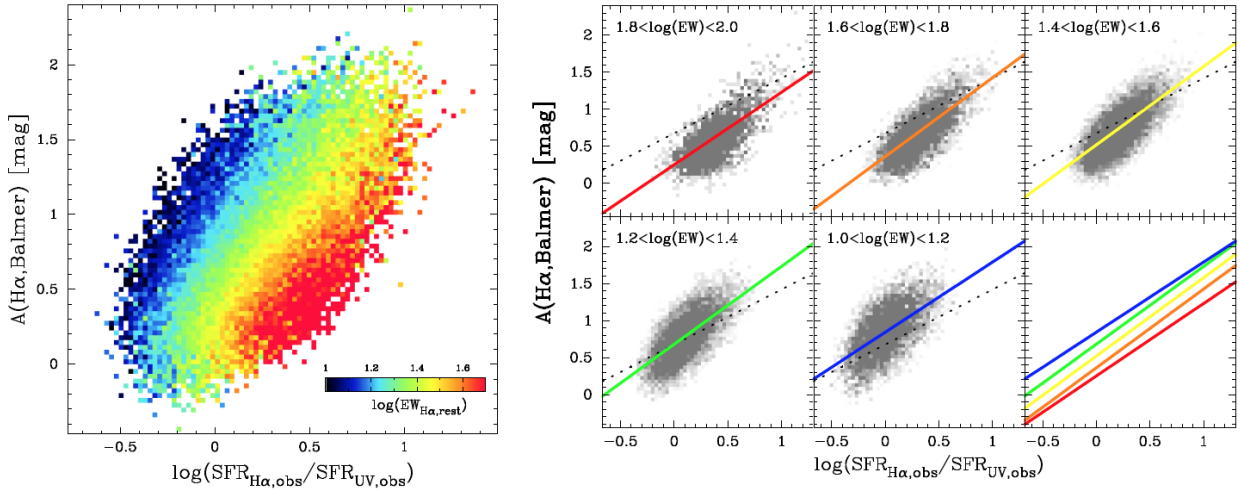} 
 \end{center}
\vspace{-3mm}
\caption{The same plot as Fig.~\ref{fig:AHa_vs_HaUV_mass}, but colour-coded based on the average EW$_{\rm H\alpha}$ at each pixel (left panel). We also show how the $A_{\rm H\alpha}$--H$\alpha$/UV correlation changes with different EW$_{\rm H\alpha}$ (right panel). Galaxies with higher EW$_{\rm H\alpha}$ (i.e.\ with younger ages) tend to be less obscured by dust at fixed H$\alpha$/UV ratio. Similar to Fig.~\ref{fig:AHa_vs_HaUV_mass}, we apply 60$\times$60 gridding to create the right-panel of this plot.}
\label{fig:AHa_vs_HaUV_EW}
\end{figure*}
The stellar mass trend reported here could partly be explained by a stellar mass dependence of the star formation history. As shown by \cite{wuyts2013}, an intrinsic $L_{\rm H\alpha}$/$L_{\rm UV}$ ratio evolves with galactic age and/or SF history of galaxies, in the sense that H$\alpha$/UV ratio rapidly declines with increasing age for those having exponentially declining SF history (see fig.~3 in \citealt{wuyts2013})\footnote{\cite{wuyts2013} show the evolution of the $L_{\rm H\alpha}$/$L_{\rm 2800}$ ratio (not $L_{\rm H\alpha}$/$L_{\rm FUV}$), but the situation is qualitatively the same. Since contribution from low-mass stars to the FUV luminosities should be smaller than NUV, the age dependence of the $L_{\rm H\alpha}$/$L_{\rm FUV}$ ratio could be even smaller than that of $L_{\rm H\alpha}$/$L_{\rm NUV}$.}. Although this effect is expected to be small, as long as the SFR does not drop very rapidly with time (see e.g.\ \citealt{hao2011}), our result may suggest that the SF history of massive galaxies are more like exponentially declining with shorter time scale, resulting in a FUV flux excess (compared with H$\alpha$) due to the contribution from lower-mass stars.

Another possibility is that the stellar mass trend seen in Fig.~\ref{fig:AHa_vs_HaUV_mass} can be explained by a stellar mass dependence of dust geometry within the galaxies. It is expected that H$\alpha$ emission originates from star-forming regions (near O-type stars) with enhanced level of dust extinction due to the effects of optically thick short-lived birth clouds. As a result, nebular emission (including H$\alpha$) suffers extra extinction correction compared to the continuum light at the same wavelength range (e.g.\ \citealt{calzetti1994}). As illustrated by \cite{price2014}, the more the SF regions are distributed uniformly over the galaxies, the difference between stellar and nebular extinction level becomes smaller. On the contrary, if stars are formed in a compact region (e.g.\ in the central core of galaxies), H$\alpha$ emission would more severely suffer from the dust extinction effects than continuum light at the same wavelength. Our result suggests that more massive galaxies tend to have higher level of extra extinction towards the nebular region ($A_{\rm nebular}$$>$$A_{\rm cont}$), whilst SF regions in low-mass systems tend to be more widely spread over the galaxies (hence $A_{\rm nebular}\sim A_{\rm cont}$). We note that this point will be further discussed in Section~4.4. 

We also comment that we cannot completely rule out the possibility that the observed trend is (partly) produced by the aperture correction effect---we recall that the H$\alpha$ fluxes are measured with a limited size of aperture. Although we believe that the effect is small, this point will be discussed more in detail in Section~4.1. 

\begin{figure*}
\vspace{1mm}
 \begin{center}
  \includegraphics[angle=0,width=16.5cm]{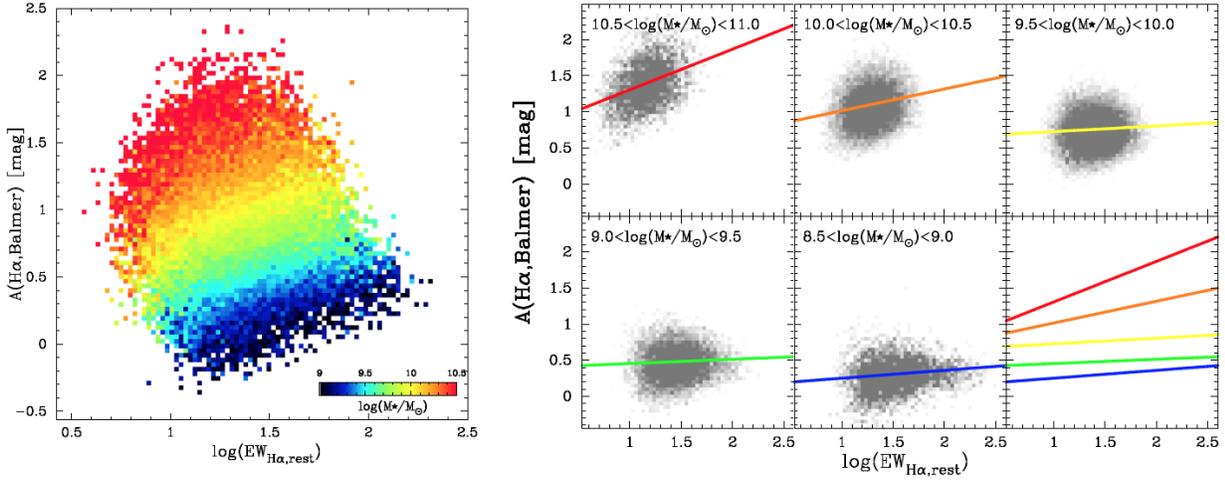} 
 \end{center}
\vspace{-3mm}
\caption{(Left): Dust extinction ($A_{\rm H\alpha}$) derived from Balmer decrement plotted against the rest-frame EW$_{\rm H\alpha}$. The colour coding indicates the stellar mass at each pixel, as we performed in the previous plots. (Right): The same plot for different stellar mass bin as indicated in the plot. It is evident that the location of the $A_{\rm H\alpha}$ versus EW$_{\rm H\alpha}$ relation is strongly dependent on stellar mass. Interestingly, the slope of the $A_{\rm H\alpha}$ versus EW$_{\rm H\alpha}$ relation changes with stellar mass: $A_{\rm H\alpha}$ increases with increasing EW$_{\rm H\alpha}$ for massive galaxies, whilst the relation is nearly flat for low-mass galaxies. }
\label{fig:AHa_vs_EW_mass}
\end{figure*}

\subsection{Dependence on EW$_{\rm H\alpha}$}

We pointed out in Fig.~\ref{fig:AHa_vs_HaUV_mass} that there remains a positive correlation between $A_{\rm H\alpha}$ and H$\alpha$/UV {\it at fixed stellar mass}. This implies that we cannot fully determine the dust extinction of galaxies with stellar mass alone. In this section, we will investigate the $A_{\rm H\alpha}$ versus H$\alpha$/UV relation as a function of another observable, EW$_{\rm H\alpha}$, which is generally more directly linked to the age of galaxies (i.e.\ younger galaxies have higher EW$_{\rm H\alpha}$). We note that EW$_{\rm H\alpha}$ indicates the {\it rest-frame} H$\alpha$ equivalent widths in the remainder of this paper.

The left panel of Fig.~\ref{fig:AHa_vs_HaUV_EW} shows the $A_{\rm H\alpha}$--H$\alpha$/UV diagram, colour-coded based on the average EW$_{\rm H\alpha}$ computed at each point (i.e.\ redder colours indicate higher EW$_{\rm H\alpha}$). As we performed in Fig.~\ref{fig:AHa_vs_HaUV_mass}, we apply 90$\times$90 gridding and compute average EW$_{\rm H\alpha}$ in each pixel, by requesting a minimum sample size of $N_{\rm pix}=4$ in each pixel. This plot demonstrates that galaxies with higher EW$_{\rm H\alpha}$ tend to show lower dust extinction at fixed H$\alpha$/UV ratio (or equivalently, galaxies with higher EW$_{\rm H\alpha}$ tend to show higher H$\alpha$/UV ratio at fixed $A_{\rm H\alpha}$), which can probably be explained by the same reason as discussed in the previous section: i.e.\ galaxies with higher EW$_{\rm H\alpha}$ are expected to have younger stellar age with little contribution from old stars to the FUV luminosities. In fact, galaxies with EW$\gsim$50\AA\ runs through the (0, 0) point on Fig.~\ref{fig:AHa_vs_HaUV_EW}, whilst low-EW galaxies show significant offsets towards higher $A_{\rm H\alpha}$. This result again suggests that a simple conversion from H$\alpha$/UV to $A_{\rm H\alpha}$ could be too simplistic. 

Following the procedure that we adopted when deriving the stellar mass dependence of the $A_{\rm H\alpha}$--H$\alpha$/UV relation in Section~3.2, we here attempt to obtain an empirical calibration for deriving $A_{\rm H\alpha}$ from EW$_{\rm H\alpha}$ and H$\alpha$/UV ratio. We assume a similar form of conversion equation as we did in Section~3.2: 
\begin{eqnarray}
A_{\rm H\alpha} = a(\log {\rm EW_{H\alpha}})\times \log ({\rm H\alpha/UV}) + b(\log {\rm EW_{H\alpha}}),
\end{eqnarray}
where $a(\log {\rm EW_{H\alpha}})= a_1\log {\rm EW_{H\alpha}} + a_2$ and $b(\log {\rm EW_{H\alpha}})= b_1\log {\rm EW_{H\alpha}} + b_2$. We fit all SDSS--GALEX sample with the above equation, and the resultant $a(\log {\rm EW_{H\alpha}})$ and $b(\log {\rm EW_{H\alpha}})$ are derived as follows:
\begin{eqnarray}
a(\log {\rm EW_{H\alpha}}) = 0.101\times \log {\rm EW_{H\alpha}} +0.872, \\
b(\log {\rm EW_{H\alpha}}) = -0.776\times \log {\rm EW_{H\alpha}} +1.688.
\end{eqnarray}
This conversion equation can be an useful tool for predicting dust extinction, particularly when stellar mass estimate is not available. Another advantage of this prescription is that it is described only with {\it observed} quantities (EW$_{\rm H\alpha}$ and H$\alpha$/UV without correction), and therefore it does not suffer from uncertainties regarding stellar mass estimates accompanied by SED fitting.  

The stellar mass dependence of the $A_{\rm H\alpha}$--H$\alpha$/UV ratio presented in Section~3.2 is based on all SF galaxies selected from SDSS sample using BPT diagram (Section~2.1). However, we want to stress that the definition of SF galaxies can be different from studies to studies: e.g.\ the sample is usually biased to higher EW$_{\rm H\alpha}$ galaxies in the case of high-$z$ studies (hence biased to higher specific-SFR galaxies with young stellar population). We will discuss later in Section~4.5 whether our new prescription can be applicable to high-redshift galaxies.

\begin{table*}
  \caption{Summary of the new recipes for predicting dust extinction proposed here and the corresponding sections in the paper.}\label{tab:recipes}
  \begin{center}
  \begin{small}
    \begin{tabular}{lcccccc}
      \hline
      Recipe & $a_1$ & $a_2$ & $b_1$ & $b_2$ & $\sigma_{\rm rms}$ [mag] & Section\\
      \hline
      (1) H$\alpha$/UV only........  & ---  &  ---   & 0.743$\pm$0.004  &  0.676$\pm$0.002  &  0.386 & \S~3.1 \\
      (2) mass + H$\alpha$/UV......  & 0.210$\pm$0.006 &  $-$1.597$\pm$0.068  & 0.493$\pm$0.003  & $-$4.121$\pm$0.026  & 0.282 & \S~3.2 \\
      (3) EW + H$\alpha$/UV........  & 0.101$\pm$0.014 &  0.872$\pm$0.020  & $-$0.776$\pm$0.006  & 1.688$\pm$0.008  & 0.338 & \S~3.3 \\
      (4) mass + EW..............   & 0.096$\pm$0.007 &  $-$0.717$\pm$0.073  & 0.538$\pm$0.011  & $-$4.745$\pm$0.104  & 0.317 & \S~3.4 \\
      \hline
    \end{tabular}
    \end{small}
  \end{center}
\end{table*}

\subsection{Predicting dust extinction without UV information}

The main aim of this paper is to establish an empirical link between the H$\alpha$/UV luminosity ratio and dust extinction properties (as presented in Section~3.1--3.3), but our analyses presented in the previous sections demonstrate that $A_{\rm H\alpha}$ is clearly dependent on stellar mass and EW$_{\rm H\alpha}$. In Figs.~\ref{fig:AHa_vs_HaUV_mass} and \ref{fig:AHa_vs_HaUV_EW}, it can be seen that the ``fixed-mass line'' and ``fixed-EW line'' show different slope. This motivates us to construct another prescription to predict $A_{\rm H\alpha}$ for a given mass and EW$_{\rm H\alpha}$. 

In the left panel of Fig.~\ref{fig:AHa_vs_EW_mass}, we plot $A_{\rm H\alpha}$ against EW$_{\rm H\alpha}$. The colour-coding indicates average stellar mass at each pixel (redder colours indicate higher stellar mass). There seems to be a general trend that $A_{\rm H\alpha}$ declines with EW$_{\rm H\alpha}$, but obviously, there is a large variation in the dust extinction properties at a given EW$_{\rm H\alpha}$. 
It is clear from Fig.~\ref{fig:AHa_vs_EW_mass} that there is an universal trend that more massive galaxies tend to be more highly obscured by dust at fixed EW$_{\rm H\alpha}$. In addition, it is interesting to note that, at fixed stellar mass, $A_{\rm H\alpha}$ sharply increases with increasing EW$_{\rm H\alpha}$ for massive galaxies, whilst this trend is not visible for low-mass galaxies ($A_{\rm H\alpha}$ always stay at $\lsim$0.5~mag regardless of EW$_{\rm H\alpha}$: see the right panel of Fig.~\ref{fig:AHa_vs_EW_mass}). This result suggests that the nature of SF galaxies is strongly dependent on stellar mass.

We can now establish an empirical prescription for deriving $A_{\rm H\alpha}$ from $M_{\star}$ and EW$_{\rm H\alpha}$. We assume a similar form of conversion equation as we did in the previous sections (but this time we do not require UV information): 
\begin{eqnarray}
A_{\rm H\alpha} = a(\log M_{\star})\times \log({\rm EW_{H\alpha}}) + b(\log M_{\star}),
\end{eqnarray}
where $a(\log M_{\star})= a_1\log M_{\star} + a_2$ and $b(\log M_{\star})= b_1\log M_{\star} + b_2$. 
We fit all SDSS--GALEX sample with the above equation, and the resultant $a(\log M_{\star})$ and $b(\log M_{\star})$ are described as follows:
\begin{eqnarray}
a(\log M_{\star}) = 0.096\times \log M_{\star} -0.717, \\
b(\log M_{\star}) = 0.538\times \log M_{\star} -4.745.
\end{eqnarray}

Our results suggest that high-mass and low-mass galaxies show different behaviour on the $A_{\rm H\alpha}$ versus EW$_{\rm H\alpha}$ diagram, probably indicating that their dust extinction properties are different. It is possible that massive galaxies tend to become more like nuclear starbursts with increasing EW$_{\rm H\alpha}$ (hence dusty starbursts with high specific-SFR), whilst low-mass galaxies tend to form stars over the galaxy discs, regardless of their specific-SFR. 

\begin{figure}
\vspace{3mm}
 \begin{center}
  \includegraphics[angle=270,width=8.0cm]{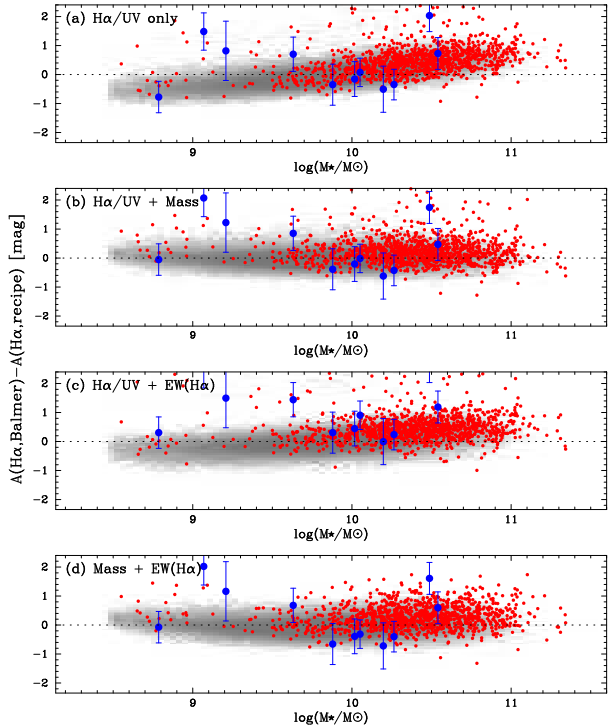} 
 \end{center}
\vspace{-1mm}
\caption{Comparison of the new recipes for predicting $A_{\rm H\alpha}$ proposed in this paper. Here we show the difference between the $A_{\rm H\alpha}$ derived from Balmer decrement and those derived using the new recipes. The top panel shows the result for the case using only H$\alpha$/UV ratio. The scatter becomes smaller in all cases (panels [b]--[d]), although they are not necessarily dramatic improvements (see also Table~1). We also show the FIR-detected galaxies (AKARI sources) with red circles, as well as $z\sim 2.5$ galaxies from \citet{shimakawa2015b} with blue circles. }
\label{fig:compare_recipe}
\end{figure}

\subsection{Summary: a new recipe for predicting dust extinction correction in the absence of Balmer decrement}

We have demonstrated that, as expected, $A_{\rm H\alpha}$ derived from Balmer decrement shows positive correlation with the observed SFR$_{\rm H\alpha}$/SFR$_{\rm UV}$ ratio (Section~3.1). This is probably a natural consequence reflecting the fact that H$\alpha$ is less sensitive to dust extinction than UV continuum light, consistent with what we had initially expected. However, we pointed out that there remains a large scatter around the $A_{\rm H\alpha}$--H$\alpha$/UV correlation, and therefore we suggest that predicting dust extinction of individual galaxies using only H$\alpha$/UV ratio could be highly uncertain. 

In this paper, we have shown that the scatter around the $A_{\rm H\alpha}$--H$\alpha$/UV correlation can be reduced by combining stellar mass and/or EW$_{\rm H\alpha}$, and we proposed new recipes for predicting dust extinction correction using a combination of H$\alpha$/UV, stellar mass, and EW$_{\rm H\alpha}$. The first approach is to use stellar mass and H$\alpha$/UV ratio (Section~3.2). The second approach is to use EW$_{\rm H\alpha}$ and H$\alpha$/UV ratio (Section~3.3), which does not require stellar mass estimate, so that it is particularly useful when multi-band photometry is not available. The final approach presented in this paper is to use stellar mass and EW$_{\rm H\alpha}$ (Section~3.4). This approach has an advantage in the absence of UV information. In Table~1, we summarize the proposed recipes outlined here. 

Overall, which is the best way to predict dust extinction among the proposed recipes? From Table~1, it can be seen that we can reduce the scatter around the best-fit relation in all cases (by $\sim$15--30\%), with the ``mass + H$\alpha$/UV'' approach being the best. Although they are not necessarily dramatic improvements, we stress that these new recipes would be able to provide more realistic dust extinction levels for individual galaxies than the simple H$\alpha$/UV approach. In Fig.~\ref{fig:compare_recipe}, we show the difference between $A_{\rm H\alpha}$ from Balmer decrement and those from our new recipes, as a function of stellar mass. In the case of simple H$\alpha$/UV method (panel-[a]), there clearly remains a systematic uncertainty: we tend to underestimate $A_{\rm H\alpha}$ for massive galaxies, while we tend to overestimate $A_{\rm H\alpha}$ for low-mass galaxies. Interestingly, for low-mass galaxies, a constant $\sim$0.5--1.0~mag correction would be more realistic than applying the H$\alpha$/UV approach (see panels for low-mass galaxies in Fig.~\ref{fig:AHa_vs_EW_mass}).

In Fig.~\ref{fig:compare_recipe}, we also plot our AKARI FIR-detected galaxies as well (red symbols). Their dust extinction level seems to be significantly underestimated by the simple H$\alpha$/UV approach (as they tend to be massive galaxies), but the situation is improved with the new recipes. An important message of this paper is that we can predict a reasonable dust extinction correction (for H$\alpha$) using UV--optical information alone, even in the absence of H$\beta$ lines or deep FIR photometry. 

\begin{figure*}
\vspace{0mm}
 \begin{center}
  \includegraphics[angle=0,width=15.2cm]{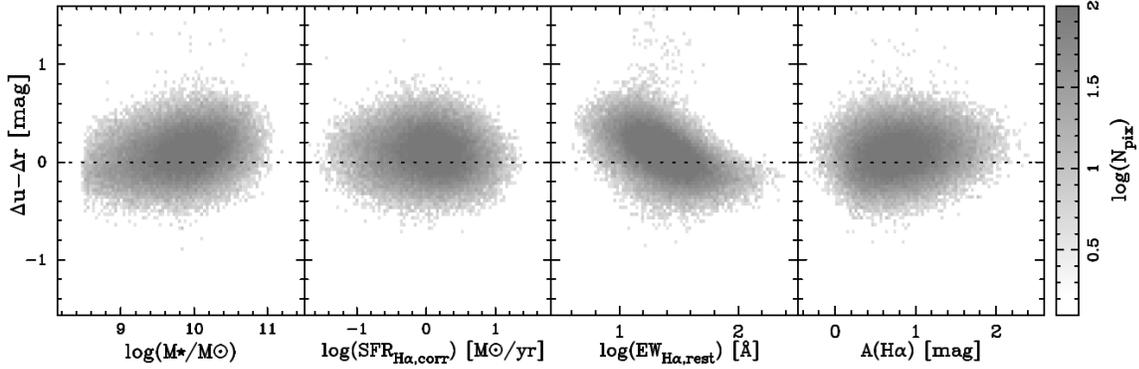} 
 \end{center}
\vspace{-2mm}
\caption{Difference of aperture correction derived from $u$-band and $r$-band ($\Delta u - \Delta r$)  plotted against stellar mass, SFR, EW$_{\rm H\alpha}$, and $A_{\rm H\alpha}$ (from left to right). This plot demonstrates that the aperture correction factors derived from $u$-band and $r$-band agree reasonably well, and therefore we expect that the aperture correction issue would not strongly bias the results. We note that the trend is most significant for EW$_{\rm H\alpha}$, indicating that $u$-band light tend to be more strongly concentrated in the central region for higher-EW galaxies. Nevertheless, we stress that our conclusion does not change even if we use the aperture correction derived from $u$-band data.}
\label{fig:aperture_corr2}
\end{figure*}

\section{Discussion}

\subsection{Systematic effects regarding aperture correction}

The most important results presented in this study is that the $A_{\rm H\alpha}$--H$\alpha$/UV relation is strongly dependent on $M_{\star}$ and EW$_{\rm H\alpha}$. A major concern when interpreting our results is the impact of aperture correction. As mentioned in Section~2, H$\alpha$ flux measurements are made with a limited size of SDSS fibre (3$''$ in diameter). We recall that the fibre can only trace $\sim$10--20\% of the total light (at $r$-band) in extreme cases (see Fig.~\ref{fig:aperture_corr1}), and so one may be worried that there remains a large uncertainty in the {\it total} H$\alpha$ flux. 

In this study, we applied aperture correction to derive H$\alpha$ total luminosity based on the difference between fibre- and total-magnitudes measured at $r$-band continuum (see Section~2.1). If massive galaxies tend to have quiescent bulge component in their central part, and if those galaxies tend to have active star formation in their outskirts, we could underestimate H$\alpha$ aperture correction: i.e.\ SFR$_{\rm H\alpha}$ could be underestimated in this case. However, we believe it is not likely the case, because our results do not change even if we use SDSS $u$-band photometry (which is expected to reflect more recent SF activity than $r$-band) to estimate aperture correction. 

In Fig.~\ref{fig:aperture_corr2}, we plot the difference between the aperture correction values measured with $u$-band and $r$-band ($\Delta u - \Delta r$) as functions of various galaxy properties. The stellar mass dependence is only $\sim$0.2-mag level at most, showing that the strong stellar mass dependence reported in Fig.~\ref{fig:AHa_vs_HaUV_mass} (Section~3.2) cannot be fully explained by the aperture correction effect. In Fig.~\ref{fig:aperture_corr2}, it can be seen that the trend is most significant for EW$_{\rm H\alpha}$: the $\Delta u - \Delta r$ value declines with increasing EW$_{\rm H\alpha}$. This would indicate that galaxies with higher EW$_{\rm H\alpha}$ tend to be forming stars in more compact regions (hence $u$-band light is more centrally concentrated), but the difference is still too small to fully explain the trend we have shown in Fig.~\ref{fig:AHa_vs_HaUV_EW} (Section~3.3). We note, however, that $u$-band is not necessarily a perfect way to trace very recent star forming activity (compared with H$\alpha$), and so we cannot completely rule out the possibility that even $u$-band light distribution might differ from that of H$\alpha$ emission. Furthermore, we need to assume in this paper that EW$_{\rm H\alpha}$ and $A_{\rm H\alpha}$ derived from fibre spectroscopy can be applicable to outer part of galaxies. This is inevitable as far as we rely on the SDSS data, and it is essential to make observations to map H$\alpha$ and H$\beta$ emission over the galaxies with large aperture, in order to completely verify our results. 

\begin{figure}
\vspace{4mm}
 \begin{center}
  \includegraphics[angle=0,width=7.5cm]{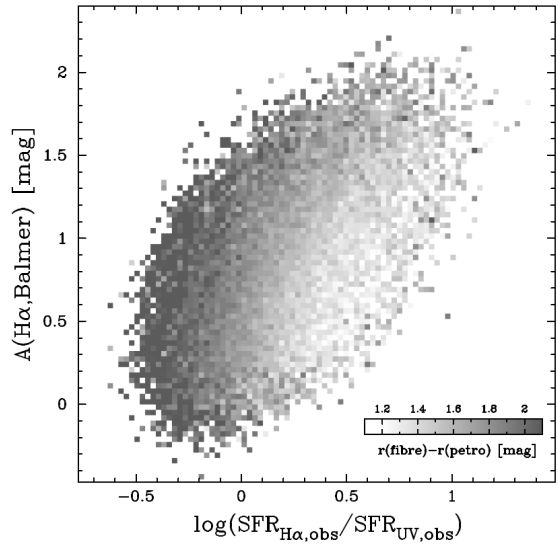} 
 \end{center}
\vspace{-1mm}
\caption{The $A_{\rm H\alpha}$ versus H$\alpha$/UV plot (same as Figs.~5--7), with the grey-scale indicating the average aperture correction value for galaxies at each pixel. Galaxies located top-left corner of this plot tend to require larger aperture correction. We recall that galaxies located at the top left corner of this plot tend to have lower EW$_{\rm H\alpha}$ (see Fig.~7). Therefore, the trend seen in this plot is actually equivalent to Figs.~2 and 7, where we showed a decreasing trend of aperture correction with EW$_{\rm H\alpha}$.}
\label{fig:AHa_vs_HaUV_apercorr}
\end{figure}

We recall that the adopted aperture correction value is correlated with EW$_{\rm H\alpha}$ as reported in Fig.~\ref{fig:aperture_corr1}: i.e.\ we need to apply {\it larger} aperture correction for {\it lower} EW$_{\rm H\alpha}$ galaxies.  We therefore caution that the levels of aperture correction also change on the $A_{\rm H\alpha}$ versus H$\alpha$/UV plot, in the sense that galaxies located at the upper envelope of the $A_{\rm H\alpha}$--H$\alpha$/UV relation tend to require larger aperture correction (Fig.~\ref{fig:AHa_vs_HaUV_apercorr}). Nevertheless, as discussed in Section~2.1, we attribute this trend to the physical reason that galaxies with higher EW$_{\rm H\alpha}$ tend to be more compact, and we believe that the trend between EW$_{\rm H\alpha}$ and the levels of aperture correction reported in Fig.~\ref{fig:AHa_vs_HaUV_apercorr} does not make any artificial bias on our conclusion.

\begin{figure*}
\vspace{2mm}
 \begin{center}
 \includegraphics[angle=0,width=16.5cm]{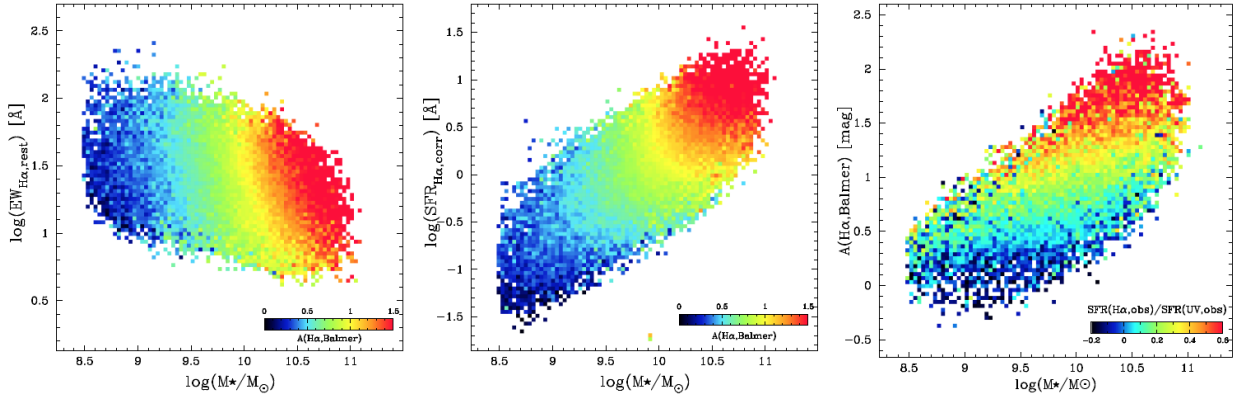} 
 \end{center}
\vspace{-1mm}
\caption{(Left): The EW$_{\rm H\alpha}$ versus stellar mass diagram, colour coded based on the dust extinction ($A_{\rm H\alpha}$). The fact that a fixed-$A_{\rm H\alpha}$ line does not run completely perpendicularly on this diagram suggests that we can provide a more reliable estimate of dust extinction by using stellar mass and EW$_{\rm H\alpha}$. (Middle): A similar plot to the left panel, but here we show the dust extinction levels of galaxies on the SFR--$M_{\star}$ diagram. This plot again shows that more massive galaxies tend to be dustier, but the behaviour is slightly more complex compared to the left-hand panel, implying that EW$_{\rm H\alpha}$ can be a better tool for estimating dust extinction of galaxies. (Right): Dust extinction versus stellar mass plot, colour coded based on the observed SFR$_{\rm H\alpha}$/SFR$_{\rm UV}$ ratio. This plot further demonstrates that predicting dust extinction from stellar mass alone can be too simplistic, whilst we can reduce the systematic uncertainty by using stellar mass {\it and} H$\alpha$/UV ratio. This is the most important finding of this work. }
\label{fig:Mstar_dependence}
\end{figure*}

\subsection{Can we predict $A_{\rm H\alpha}$ with stellar mass alone?}

Recent studies suggest that stellar mass is the most important parameter which determines the dust extinction levels for SF galaxies (e.g.\ \citealt{garn2010b}). Using our own datasets, we also confirm that the scatter around the best-fit $A_{\rm H\alpha}$--stellar mass relation is fairly small (0.32~mag): this is actually even smaller than the H$\alpha$/UV approach ($\sim$0.38~mag), and is comparable to those achieved with our new recipe (see Table~1). Furthermore, the correlation between dust extinction and stellar mass is reported to be unchanged at least out to $z\sim 1.5$ (\citealt{garn2010a}; \citealt{sobral2012}; \citealt{ibar2013}; \citealt{dominguez2013}), and so stellar mass can be an useful indicator for a rough estimate of dust extinction of galaxies at all redshifts in the absence of the Balmer decrement or deep FIR photometry (but see \citealt{oteo2014}; \citealt{pannella2015}; \citealt{shimakawa2015a}). However, as demonstrated in this work, it might be too simplistic to blindly apply the $M_{\star}$-dependent dust extinction correction. In this section, we discuss how well we can really predict dust extinction from stellar mass alone.  
 
In the left- and middle-panel of Fig.~\ref{fig:Mstar_dependence}, we plot EW$_{\rm H\alpha}$ and SFR as a function of stellar mass. The colour-coding indicates the average $A_{\rm H\alpha}$ derived from Balmer decrement at each point (redder colours indicate higher dust extinction). The left panel is essentially equivalent to Fig.~\ref{fig:AHa_vs_EW_mass}, where we showed the stellar mass dependence of the $A_{\rm H\alpha}$--EW$_{\rm H\alpha}$ relation. This plot further demonstrates that more massive galaxies tend to be dustier on average (consistent with many other studies; e.g.\ \citealt{whitaker2012}), but it is also important to note that, at fixed $M_{\star}$, galaxies with higher EW$_{\rm H\alpha}$ tend to be more highly obscured by dust. On the other hand, the middle panel of Fig.~\ref{fig:Mstar_dependence} shows how the average dust extinction of galaxies changes on the SFR--$M_{\star}$ diagram. Again, more massive galaxies tend to be dustier, whilst the behaviour of a fixed-$A_{\rm H\alpha}$ line on the SFR--$M_{\star}$ diagram is rather more complex compared with the EW--$M_{\star}$ plot (left panel). It is interesting to note that a more simple observable (EW$_{\rm H\alpha}$) can be a better tool for predicting dust extinction of galaxies---and this is why we exploit EW$_{\rm H\alpha}$ (a proxy for the age of stellar population) rather than SFRs when we establish an empirical calibration in Section~3. 

Finally, in the right panel of Fig.~\ref{fig:Mstar_dependence}, we show $A_{\rm H\alpha}$ against stellar mass. Here we apply colour-coding based on the H$\alpha$/UV ratio as we investigated in this paper. This plot is equivalent to Fig.~\ref{fig:AHa_vs_HaUV_mass}, but more directly demonstrates our important result that a simple stellar-mass dependent dust extinction correction is not a perfect way: by adding H$\alpha$/UV ratio we can have a more realistic dust extinction correction in the absence of Balmer decrement and/or deep FIR photometry. 

\begin{figure}
\vspace{2mm}
 \begin{center}
  \includegraphics[angle=0,width=7.5cm]{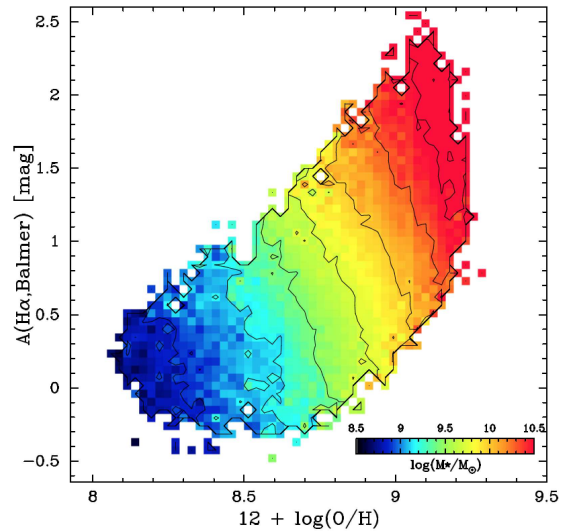}
 \end{center}
\vspace{-3mm}
\caption{Dust extinction versus gas-phase metallicity for our SDSS--GALEX star-forming galaxy sample, colour-coded based on the average stellar mass at each point. We applied 60$\times$60 gridding for making this plot. There is a general trend that the dust extinction level increases with increasing metallicity. It is also notable that the dust extinction increases with stellar mass at fixed metallicity, and at the same time, metallicity increases with stellar mass at fixed $A_{\rm H\alpha}$. }
\label{fig:AHa_vs_Metal}
\end{figure}
\begin{figure*}
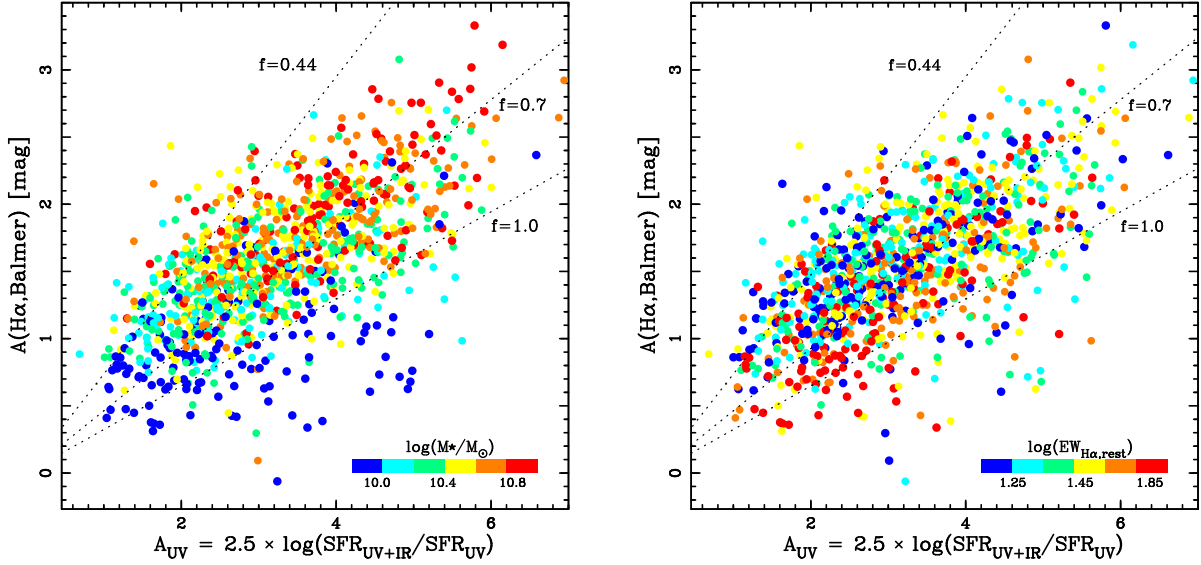

\vspace{2mm}
 \begin{center}
  \includegraphics[angle=270,width=7.5cm]{fig14a.ps} 
  \hspace{8mm}\includegraphics[angle=270,width=7.5cm]{fig14b.ps} 
 \end{center}
\vspace{0mm}
\caption{Comparison between the H$\alpha$ (nebular) dust extinction from Balmer decrement and UV (continuum) dust extinction measured with IR/UV ratio, for AKARI FIS detected galaxies. The colour coding of the data points represent the stellar mass (left panel) and the rest-frame EW$_{\rm H\alpha}$ (right panel): i.e.\ redder colours indicate higher $M_{\star}$ and higher EW$_{\rm H\alpha}$ in the left and right panel, respectively. The three dotted lines correspond to the extra extinction factor of $f=0.44$, $0.7$, and $1.0$ (in the case of Calzetti extinction law). This plot demonstrates that more massive galaxies and/or lower EW$_{\rm H\alpha}$ galaxies tend to have higher extra extinction towards the nebular regions (i.e.\ smaller $f$ value). }
\label{fig:AHa_vs_Auv}
\end{figure*}

\subsection{Metallicity dependence}

Dust is formed by metals, and so the dust extinction may be correlated with metallicity of galaxies (e.g.\ \citealt{xiao2012}). In Fig.~\ref{fig:AHa_vs_Metal}, we plot H$\alpha$ dust extinction derived by Balmer decrement for our SDSS--GALEX star-forming galaxies against their (gas-phase) metallicity derived by \cite{tremonti2004}. The colour coding of this plot indicates average stellar mass at each point (redder colours indicate higher $M_{\star}$). 

It is clear from Fig.~\ref{fig:AHa_vs_Metal} that there is an overall trend that dust extinction level increases with increasing metallicity. It is also evident that both dust extinction and metallicity increase with increasing stellar mass. At fixed metallicity, dust extinction level slightly increases with stellar mass. Similarly, at fixed $A_{\rm H\alpha}$, metallicity also increases with stellar mass. At fixed stellar mass, on the other hand, it looks as if dust extinction level sharply {\it decreases} with increasing metallicity. This may sound surprising, because the trend is completely opposite to the overall dust--metallicity correlation. We note, however, that this is just an {\it average} behaviour---we find that individual data points at fixed stellar mass do not align so tightly on the fixed-mass line in Fig.~\ref{fig:AHa_vs_Metal}. Neveretheless, it is interesting to point out that a small variation (scatter) around the mass--metallicity relation is related to the variation of dust properties of galaxies at fixed stellar mass (see consistent results for high-$z$ galaxies by \citealt{zahid2014}). 

Detailed modeling of the chemical evolution and dust formation/destruction processes within galaxies is beyond the scope of this paper, but we comment that our results are basically unchanged even if we replace the stellar mass parameter with the metallicity. However, we recall that the main aim of this paper is to establish an empirical link between dust extinction properties of galaxies and some quantities which can easily be obtained. Deriving metallicity of galaxies is usually a more complex issue (and it requires detection of several faint emission lines), hence we decided to use stellar mass (rather than metallicity) when we derive the empirical calibration in Section~3.

\subsection{Stellar mass and EW$_{\rm H\alpha}$ dependence of the extra extinction towards nebular regions}

It is suggested that there is a need for an extra extinction correction to nebular emission lines (including H$\alpha$) compared with continuum light at the same wavelength. This is because nebular emission lines are originated in the dust-rich birth cloud, whilst other galaxy continuum light are only affected by diffuse interstellar dust (see e.g.\ \citealt{calzetti1997}; \citealt{charlot2000}; \citealt{cidfernandes2005}; \citealt{wuyts2011}; \citealt{steidel2014}). Using a sample of local star-forming galaxies, \cite{calzetti2000} argue that the colour excess of the stellar continuum, $E_{\rm s}(B-V)$, is linked to that of nebular emission lines, $E_{\rm gas}(B-V)$, with $E_{\rm s}(B-V) = 0.44 \times E_{\rm gas}(B-V)$ (i.e.\ extra extinction factor, $f=0.44$). In fact, the amount of this extra extinction factor ($f$) is still under debate. For instance, \cite{wuyts2013} show a polynomial function to derive $A_{\rm extra}$ from the continuum extinction ($A_{\rm cont}$), hence the $f$ value may not necessarily be a constant value for all types of galaxies. It is also suggested by recent studies that the typical $f$ value may be higher ($f=0.5-1.0$) for high-redshift galaxies (see e.g. \citealt{erb2006}; \citealt{reddy2010}; \citealt{kashino2013}; \citealt{valentino2015}). 

Taking advantage of the AKARI FIR photometry (which captures the peak of dust emission), we compare in Fig.~\ref{fig:AHa_vs_Auv} the dust extinction derived from Balmer decrement ($A_{\rm H\alpha}$; i.e.\ nebular extinction) and the IR/UV ratio ($A_{\rm UV}$; i.e.\ continuum extinction). It can be seen that there is broadly a positive correlation between the two independent measurements of dust extinction properties, but the left panel of Fig.~\ref{fig:AHa_vs_Auv} also demonstrates that more massive galaxies tend to have higher $A_{\rm H\alpha}$ than low-mass galaxies at fixed $A_{\rm UV}$. Similarly, galaxies with lower EW$_{\rm H\alpha}$ tend to have higher $A_{\rm H\alpha}$ at fixed $A_{\rm UV}$ (see right panel of Fig.~\ref{fig:AHa_vs_Auv}).

In Fig.~\ref{fig:AHa_vs_Auv}, we show three dotted lines corresponding to $f=$0.44, 0.7, 1.0, after properly converting $A_{\rm UV}$ to the continuum extinction at H$\alpha$. We find that most of our galaxy sample (AKARI-detected star-forming galaxies) are distributed in the range between $f$$=$0.44 and $f$$=$1.0, and at the same time, we also find an interesting trend that more massive galaxies tend to have smaller $f$ value: i.e. higher extinction towards the nebular regions. Interestingly, a very recent study by \cite{puglisi2015} made a similar suggestion for high-redshift galaxies (see their fig.~10). We therefore suggest that the stellar mass (or EW$_{\rm H\alpha}$) trend of $f$ value reported in our current study may hold over the cosmic time. Assuming that the difference in $f$ value is due to dust geometry within the galaxies (\citealt{price2014}), our results suggest that high-$z$ SF galaxies, which are often reported to have $f\sim 1$, are more like low-mass galaxies in the low-redshift Universe, which tend to be forming stars over the disc. 

We caution that the stellar mass and/or EW$_{\rm H\alpha}$ dependence of the $f$ value reported here is implicitly incorporated when establishing our empirical recipes for predicting dust extinction levels of galaxies in Section~3. For example, for massive galaxies, our H$\alpha$/UV methods automatically assume {\it small} $f$ values. This might be misleading, particularly if high-redshift galaxies tend to have {\it larger} $f$ value comparable to low-mass galaxies in the local Universe---in this case we might {\it over-estimate} the dust extinction levels of (massive) high-$z$ galaxies if we simply applied our local calibration. However, unfortunately, high-$z$ galaxy sample currently available is too small to test this possibility (see blue symbols in Fig.~\ref{fig:compare_recipe}). Also, by considering the situation that the typical "$f$" value reported for high-$z$ galaxies significantly vary from studies to studies, it would be impossible at this moment to conclude the above possibility (see also Section~4.5).

We note that we have used \cite{calzetti2000} extinction law to convert $A_{\rm UV}$ to the (continuum) extinction level at H$\alpha$ wavelength, to be consistent with all the analyses presented in this paper. If we instead used other extinction curves, then the absolute $f$ values would change by a factor of $\sim$2$\times$ at maximum: e.g.\ more specifically, the slope of the dotted lines on Fig.~\ref{fig:AHa_vs_Auv} would be reduced by $\sim$10\% when we apply the Milky-way dust extinction curve of \cite{cardelli1989}, and by $\sim$50\% in the case of the SMC-type extinction curve of \cite{gordon2003}. Nevertheless, the {\it relative} difference in $f$ values between high-mass and low-mass galaxies does not change as long as we assume the same type of extinction curve for all galaxies in the sample. We note that this assumption itself is uncertain, and the shape of extinction curves may depend on galaxy properties such as (specific) SFRs (see e.g.\ \citealt{wild2011}). However, unfortunately, it is impossible to assess the shape of extinction curve (and its stellar mass dependence) with the current dataset alone.

\subsection{Application to high-redshift galaxies}

The main aim of this study is to construct an empirical calibration to predict dust extinction from the observed H$\alpha$/UV ratio (combined with some other galaxy properties). The new recipes developed in this paper should be an useful tool particularly when Balmer decrement or deep FIR photometry is not available, which is usually the case for high-redshift studies. In this final subsection, we discuss if our new recipes can really be applicable to high-redshift galaxies.

Although it is becoming easier to have good quality NIR spectra for high-$z$ galaxies, it is still very challenging to detect H$\beta$ lines for individual galaxies even with 8-m class telescope. We use $z=2.5$ galaxy sample from \cite{shimakawa2015b}, for which H$\alpha$ and H$\beta$ fluxes are both directly measured with deep spectroscopy with MOSFIRE on Keck. In Fig.~\ref{fig:compare_recipe}, we plot our high-$z$ galaxy sample with the blue circles on top of the distribution of our local galaxy sample. Obviously, our high-$z$ sample currently available is by far too small to tell if our procedure works well for high-$z$ galaxies. Therefore, we stress that the applicability of our new recipe to high-redshift galaxies needs to be tested using larger high-$z$ galaxy sample with high S/N spectra in the future. A growing number of extensive NIR spectroscopic observational campaign with new facilities are now under way (e.g.\ \citealt{steidel2014}; \citealt{reddy2015}), and also, future thirty-meter class telescopes will of course deliver high-quality spectra for high-redshift galaxies. We believe that those new observations will revolutionise the situation and make it easier to access Balmer decrement for high-$z$ galaxies, allowing us to test if our new empirical recipes for predicting dust extinction properties established in the local Universe really work for high-redshift galaxies.

\section{Summary}

We presented an empirical calibration between the observed H$\alpha$-to-FUV ratio and dust extinction ($A_{\rm H\alpha}$ derived from Balmer decrement) for SF galaxies at $0.02<z<0.1$ (selected with BPT diagram) using our SDSS(DR7)--GALEX(GR5) matched galaxy sample, helped by AKARI FIR all-sky data. We confirmed that $A_{\rm H\alpha}$ increases with increasing H$\alpha$/UV ratio (as expected), but there exists a considerable scatter around the $A_{\rm H\alpha}$--H$\alpha$/UV relation. 

An important finding of this study is that the scatter around the $A_{\rm H\alpha}$--H$\alpha$/UV relation is largely dependent on stellar mass, as well as EW$_{\rm H\alpha}$ of galaxies. At fixed H$\alpha$/UV ratio, galaxies with higher stellar mass or lower EW$_{\rm H\alpha}$ tend to have higher $A_{\rm H\alpha}$, which could be explained by the stellar mass dependence (and/or EW$_{\rm H\alpha}$ dependence) of the SF history: an intrinsic H$\alpha$/UV luminosity ratio is expected to rapidly decline with galactic age, particularly when assuming exponentially declining SFR (\citealt{wuyts2013}). Another possibility is that more massive galaxies tend to have higher extra extinction towards nebular regions, so that H$\alpha$ is more heavily attenuated with respect to UV continuum light (see below).

At fixed stellar mass, $A_{\rm H\alpha}$ is still positively correlated with H$\alpha$/UV ratio and with EW$_{\rm H\alpha}$, suggesting that predicting dust extinction from stellar mass alone could be too simplistic. We have established an empirical calibration for predicting $A_{\rm H\alpha}$ using (1) stellar mass + H$\alpha$/UV ratio, (2) EW$_{\rm H\alpha}$ + H$\alpha$/UV ratio, and (3) stellar mass + EW$_{\rm H\alpha}$. By comparing the $A_{\rm H\alpha}$ value derived by Balmer decrement and those from the newly proposed methods, we find that our new methods work reasonably well (and successfully reduce the scatter around the best-fitted relation by $\sim$15--30\%), with the ``mass + H$\alpha$/UV'' method being the best. Nevertheless, the other methods can also be useful in any future studies, particularly when stellar mass estimates and/or UV continuum information are not available.

We find that $A_{\rm H\alpha}$ for high-mass galaxies increases with increasing EW$_{\rm H\alpha}$, whilst dust extinction does not increase with EW$_{\rm H\alpha}$ for low-mass systems, suggesting a different nature of SF activity for galaxies with different mass. We use AKARI FIR data to derive UV extinction ($A_{\rm UV}$) using IR/UV ratio, and test the stellar mass and EW$_{\rm H\alpha}$ dependence of the extra extinction correction factor towards the nebular region ($f$ value, defined as $E_{\rm star}(B-V) = f \times E_{\rm gas}(B-V)$). We find an interesting hint that the $f$ value is dependent on stellar mass or EW$_{\rm H\alpha}$, with more massive galaxies or low EW$_{\rm H\alpha}$ galaxies having higher extra extinction towards nebular regions (i.e.\ smaller $f$ value). Considering recent studies claiming higher $f$-value for high-$z$ galaxies, we argue that the dust geometry within high-$z$ SF galaxies resemble more like low-mass galaxies in the nearby Universe. 

An important caveat is that our empirical recipes for predicting dust extinction from H$\alpha$/UV ratio implicitly incorporate this stellar mass and/or EW$_{\rm H\alpha}$ dependence of the $f$ value: e.g.\ for massive galaxies our recipes automatically assume {\it small} $f$ value. If high-redshift galaxies tend to have {\it larger} $f$ value than local counterparts with the same mass as often reported by recent studies, our empirical recipes established with local galaxy sample might {\it over-estimate} the dust extinction levels for high-redshift galaxies. However, the current high-$z$ galaxy sample (with measured H$\alpha$/H$\beta$ ratio) is too small to test if our new recipes can be applicable to high-$z$ galaxies.

\section*{Acknowledgment}
We thank the referee for reviewing our paper and providing us with very helpful comments which improved the paper. This work was financially supported in part by a Grant-in-Aid for the Scientific Research (Nos.\,26800107; 24244015) by the Japanese Ministry of Education, Culture, Sports and Science. This research is based on observations with AKARI, a JAXA project with the participation of ESA. This research made use of the ``K-corrections calculator'' service available at http://kcor.sai.msu.ru/. M.H. and R.S. acknowledge support from the Japan Society for the Promotion of Science (JSPS) through JSPS research fellowships for Young Scientists.



\end{document}